\documentclass{elsart}

\usepackage{graphicx}

\usepackage{amssymb}

\begin{document}

\begin{frontmatter}



\title{Experimental study of high energy electron interactions in a superconducting aluminum alloy resonant bar}


\author[label6,label7] {M.~Barucci}
\author[label2,label3] {M.~Bassan}
\author[label1] {B.~Buonomo}
\author[label4] {G.~Cavallari}
\author[label2,label3]{E.~Coccia}
\author[label2] {S.~D'Antonio}
\author[label2,label3] {V.~Fafone}
\author[label1] {C.~Ligi}
\author[label6,label7] {L.~Lolli}
\author[label1] {A.~Marini}
\author[label1] {G.~Mazzitelli}
\author[label1] {G.~Modestino}
\author[label3,label1] {G.~Pizzella}
\author[label1] {L.~Quintieri}
\author[label6,label7] {L.~Risegari\thanksref{label8}}
\author[label2] {A.~Rocchi}
\author[label1] {F.~Ronga}
\author[label5] {P.~Valente}
\author[label6,label7] {G.~Ventura}
\author[label1] {S.M.~Vinko\thanksref{label9}}

\address[label1] {INFN  (Istituto Nazionale di Fisica Nucleare) Laboratori Nazionali di Frascati, I 00044 Frascati, Italy}
\address[label2] {INFN  Sezione Roma2, I 00133 Rome, Italy}
\address[label3] {Dipartimento di Fisica, Universit\`a di Tor Vergata, I  00133 Rome, Italy}
\address[label4] {CERN, CH1211 Geneva, Switzerland}
\address[label5] {INFN  Sezione di Roma1, I 00185 Rome, Italy}
\address[label6] {INFN  Sezione di Firenze, I 00185 Sesto Fiorentino, Florence, Italy}
\address[label7] {Dipartimento di Fisica, Universit\`a di Firenze, I 00185 Sesto Fiorentino, Florence, Italy}

\thanks[label8]{Present address: CSNSM, 91405 Orsay Campus, France}
\thanks[label9]{Present address: Dept. of Physics, Univ. of Oxford, Oxford OX1 3PU, UK}

\begin{abstract}
Peak amplitude measurements of the fundamental mode of oscillation of a suspended aluminum alloy bar hit by an electron beam show that the amplitude is enhanced by a factor $\sim 3.5$ when the material is in the superconducting state. This result is consistent with the cosmic ray observations made by the resonant gravitational wave detector NAUTILUS, made of the same alloy, when operated in the superconducting state. A comparison of the experimental data with the predictions of the model describing the underlying physical process is also presented.
\end{abstract}


\end{frontmatter}

\section{Introduction}
\label{intro}
In a pioneering experiment \cite{beron}  B.L.~Baron and R.~Hofstadter measured mechanical oscillations in piezoelectric disks when penetrating high energy electron beams impinged on the disks.
The authors outlined the possibility that cosmic ray  events could excite mechanical vibrations in a metallic cylinder at its resonant frequency and that they could represent a background for experiments aimed at the detection of gravitational waves (gw). The gw resonant detector NAUTILUS,
a massive (2.3 t) suspended cylinder made of an aluminum alloy (Al5056) that can be cooled down to the thermodynamic temperature of 0.1 K, has been equipped with a cosmic ray detector to study the interactions due to cosmic rays and to provide a veto against the induced events in the antenna. The results on the cosmic ray observations made by NAUTILUS can be summarized as follows: 1) when the antenna was operated at a temperature $T=0.14\ \rm{K}$, well below the transition temperature from normal-conducting ($n$) to superconducting ($s$) states of the material, the rate of high energy signals due to cosmic ray showers was larger  than the expectations based on the model describing the underlying physical processes \cite{naut1,naut2}; 2) there was no evidence of this feature when the antenna was operated at $T=1.5\ \rm{K}$, well above the transition temperature \cite{naut3}. From one side the hypothesis that this behavior was linked to the conducting state of the antenna and on the other side the incomplete knowledge at very low temperature of the thermophysical and thermodynamic parameters needed by the model have motivated an experiment (RAP) to measure the longitudinal oscillations of suspended cylindrical bars exposed to electron beam pulses of controlled energy and intensity. The experiment, performed at the Beam Test Facility (BTF) \cite{mazbtf} of the DAFNE $\Phi$-factory complex in the INFN Frascati Laboratory, has already obtained the following results: 1) the measurements  over a wide temperature interval $(4.5\ \rm{K} \leq T \leq 264\ \rm{K})$ on a bar made of the same aluminum alloy as NAUTILUS have confirmed with  good precision the validity of the model \cite{buonomo}; 2) the measurements on a pure niobium bar operated in the $n$ and $s$ state have demonstrated that the oscillation amplitude of the bar induced by the interaction with the beam depends on the state of conduction of the material \cite{bassan}.
\par\noindent In this letter we report on the measurements made on the aluminum alloy bar above and below the temperature of transition between the $s$ and $n$ state. In particular, we present a  description of the model (Section \ref{tam}),  a summary of the experimental setup (Section \ref{exset}), the collected data and analysis (Section \ref{meas}) and the comparison between the data and the model (Section \ref{comp}).

\section{Discussion of the thermo-acoustic effects}
\label{tam}
A pressure pulse is generated in a suspended cylindrical bar in the $n$ state following the interactions of an elementary particle with the bulk. This sonic pulse, due to the local thermal expansion caused by the warming up, related to the energy lost by the particle crossing the material, determines the excitation of the vibrational modes of the bar. In the experiment of Ref. \cite{strini} an aluminum bar was exposed to a proton beam and the theoretical expectations were based on a model in which the ``amplitude of the fundamental longitudinal mode of oscillation'', hereafter referred to as Amplitude, is given by:
\begin{equation}
B_{0}=\frac{2 \alpha L W}{\pi c_V M} 
\label{b0}
\end{equation}
\noindent for a beam hitting the center of the cylinder generatrix. In the previous relation $L$,  $M$ are respectively  length, mass of the cylinder, $W$ is the total energy loss of the beam in the bar,  $\alpha$ is the linear thermal expansion coefficient and  $c_V$ is the isochoric specific heat. The ratio of the thermophysical quantities $\alpha$ and $c_V$ is part of the definition of the Gr\"{u}neisen parameter of the material:
\begin{equation}
\gamma=\frac{\beta K_{T}}{\rho c_{V}}\ ,
\label{gru}
\end{equation}
\noindent where $\beta$ is the volumetric thermal expansion coefficient  ($\beta=3\alpha$ for aluminum), $K_{T}$ is the isothermal bulk modulus and $\rho$ is the mass density. The parameter $\gamma$ is a very slowly varying function of the temperature when the material is in  the $n$ state. Solution (\ref{b0}) is a particular case of a more general treatment of the problem, which includes  the paths  of the interacting particles in the bulk other than the coordinate of the impact point  \cite{cabibbo,deru,bar}. By the introduction of a vector  field  $\bold{u}(x,t)$ describing the local displacements from equilibrium, the amplitude of the mode $k$ of the cylinder oscillation is proportional to:
\begin{eqnarray}
g{_k}{^{therm}}& = & {\frac{\Delta P^{therm}}{\rho}} {\mathcal A'} {\mathcal I}{_k} \nonumber \\
                          & =  & {\frac{\gamma}{\rho}}\left| {\frac{dW}{dx}}\right|  {\mathcal I}{_k} \ ,
\label{gtherm}
\end{eqnarray}
\noindent where  $\Delta P^{therm}$ is the pressure pulse due to the sonic source previously described, $dW/dx$ is the specific energy loss of the interacting particle, $\mathcal A'$ is the cross section of the tubular zone centered on the particle path in which the effects are generated and ${\mathcal I}{_k}=\int{dl (\nabla\cdot\bold{u}{_k}(x))}$ is a line integral over the particle path involving the normal mode of oscillation $\bold{u}{_k}(x)$. The Amplitude, as given by (\ref{b0}), can be obtained starting from (\ref{gtherm}) for a thin bar $(R/L\ll 1$, where $R$ is the bar radius) and for particles hitting the central section.

\noindent In the following, for the material in $n$ state, we will compare the measured values of Amplitude to the expected value:
   \begin{equation}
    X_{therm}=B_{0}(1+\epsilon)  \ ,
    \label{xth}
    \end{equation}
where  $\epsilon$ is a corrective parameter estimated by a Monte Carlo (MC) simulation~\cite{buonomo}, which takes into account the solutions $O[(R/L)^2]$ for the modes of oscillation of a  cylinder, the transverse dimension of the beam at the impact point  and the trajectories of the secondary particles
generated in the bar. The value of  $\epsilon$ for the aluminum alloy bar used in the experiment is estimated by MC to be -0.04.

When the material is in the $s$ state, an additional sonic source could be due to the local $s$-$n$ transitions in zones centered around the interacting particle path \cite{cabibbo,deru}. The additional contribution to the amplitude of the cylinder oscillation mode $k$ is proportional to:
\begin{eqnarray*}
g{_k}{^{trans}}& = & {\frac{\Delta P^{trans}}{\rho}} {\mathcal A''} {\mathcal I}{_k}  \\
       & =  & {\frac{\gamma}{\rho}}\left[{ K_{T}\frac{\Delta V}{V}+\gamma T \frac{\Delta \mathcal{S}}{V}}\right]  {\mathcal A''}{\mathcal I}{_k} \ , 
\end{eqnarray*}
\noindent  where $\Delta V$ and $\Delta \mathcal{S}$ are the differences of the volume and entropy in the two states of conduction, while $\mathcal A''$ is the cross section of the tubular zone centered on the interacting particle path and  switched from $s$ to $n$ state, which is given  by   ${\mathcal{A''}}=(dW/dx)/(\Delta {\mathcal H}/V)$~\cite{sherman,strehl} involving the difference of enthalpy, ${\mathcal H}$, among the two states. The differences can be expressed in terms of the thermodynamic critical field
$H_c$ and it follows, in first approximation, that \cite{hake,cgs}: $\Delta V/V$=$(V_{n}-V_{s})/V$=$H_{c}(\partial H_{c}/\partial P)/(4 \pi)$ and $\Delta {\mathcal S}/V$=$( {\mathcal S}_{n}- {\mathcal S}_{s})/V$=$-H_{c}(\partial H_{c}/\partial T)/(4\pi)$. Moreover, by using the difference ($\Delta {\mathcal G}/V$=$( {\mathcal G}_{n}- {\mathcal G}_{s})/V$=$H_{c}^{2}/(8\pi)$) of the Gibbs free energy among the two states and by making the hypothesis that $H_c$ has the parabolic behavior $H_c(t)=H_c(0)(1-t^2)$, where $t=T/T_c$ and $T_c$ is the transition temperature, it follows that $\Delta {\mathcal H}/V = H_{c}^{2}(0)(1-t^2)(1+3t^2)/(8\pi)$.  In order to compare the observed data with the model predictions, we will use the ratio $\mathcal R$ of the contributions to the Amplitude due to local transition effects $(X_{trans})$ and to thermal effects in the $n$ state $(X_{therm})$. $\mathcal R$ can be expressed as:
\begin{eqnarray}
{\mathcal R} & = & \frac{X_{trans}}{X_{therm}} = \frac{ g{_0}{^{trans}}}{g{_0}{^{therm}}}  \nonumber \\
       & =  & \left[{ \frac{K_{T}}{\gamma}\frac{\Delta V}{V}+T \frac{\Delta \mathcal{S}}{V}}\right] 
       \left[{ \frac{\Delta\mathcal H}{V}}\right]^{-1} 
        \ ,
\label{r}
\end{eqnarray}
\noindent due to the existent proportionality between the mode amplitude and $g$. 

In an alternative scenario, which takes into account that local transitions do not occur, the Amplitude expected values  are given by the relation~(\ref{xth}) making use of the $\alpha$ and $c_V$ values for the $s$ state.

 \section{Experimental setup}
\label{exset}
The  experiment setup has been fully described in Ref.~\cite{buonomo}. Here we briefly recall that the test mass is a  cylindrical bar ($R=0.091\ \rm{m}$, $L=0.5\ \rm{m}$, $M=34.1\ \rm{kg}$) made of Al5056, the same aluminum alloy (nominal composition 5.2 w\% Mg and 0.1 w\% of both Cr and Mn) used for NAUTILUS. The bar hangs from the cryostat top by means of a multi-stage suspension system insuring an attenuation on the external mechanical noise of  -150 dB in the 1700-6500 Hz frequency window. The frequency of the fundamental longitudinal mode of oscillation of the bar is $f_0=5413.6\ \rm{Hz}$ below $T = 4\ \rm{K}$.  The cryostat is equipped with a $\rm{{^3}He}$  refrigerator, capable of cooling the bar down to $T\sim 0.5\ \rm{K}$. The temperatures are measured inside the cryostat by 10 thermometers controlled by a multi-channel resistance bridge. In particular, a calibrated $\rm{RuO_2}$ resistor detects the temperature of one of the bar end faces with an accuracy of  0.01 K for $T\lesssim 4\ \rm{K}$. Two piezoelectric ceramics (Pz), electrically connected in parallel, are inserted in a slot cut in the position opposite to the bar suspension point and are squeezed when the bar shrinks. In this Pz arrangement the strain measured at the bar center is proportional to the displacement of the bar end faces. The Pz output is first amplified and then sampled at 100 kHz by an ADC embedded in a VME system, hosting the data acquisition system.  The measurement of the Pz  conversion factor $\lambda$, relating voltage to oscillation amplitude, is accomplished according to a procedure based on the injection in the Pz of a sinusoidal waveform of known amplitude, with frequency $f_0$ and time duration less than  the decay time of the mechanical excitations and on the subsequent measurement of Amplitude. The procedure is correct if $R/L \ll 1$ and  a 6\% systematic error in the determination of $\lambda$ was found. Amplitude  is measured according to  $X = V_0^{\mathrm{meas}}/(G\lambda)$, where $G$  is the amplifier  gain and $V_0^{\mathrm{meas}}$ is the maximum of the signal component at frequency $f_0$, which is obtained by Fast Fourier Transform algorithms applied to the digitized Pz signals. The sign of Amplitude is taken positive or negative according to the sign of the first sampling  above the noise in the waveform generated by the Pz and sampled by the ADC. 
\noindent BTF delivers to the bar single  pulses of $\sim10\ $ns duration,  containing $N_e$ electrons of $510\pm 2$ MeV energy.  $N_e$ ranges from about $5\times 10^7$ to $10^9$ and is measured with an accuracy of $\sim 3 \%$  (for $N_e > 5\times 10^8$) by an integrating current transformer placed close to the beam exit point.  MC, already introduced in Section~\ref{tam}, estimates an average energy lost $\langle {\Delta  E}\rangle \pm \  \sigma_{\Delta  E}=195.2 \pm 70.6 \ \rm{MeV}$ for a 512 MeV electron interacting in the bar and, consequently, the total energy loss per beam pulse is given by $W=N{_e} \langle {\Delta  E}\rangle \  , \ \sigma_{W}=\sqrt{N{_e}}\  \sigma_{\Delta E}$.

\section{Measurements and data analysis}
\label{meas}
\begin{figure}[htbp]
\begin{center}
\includegraphics[width=1.0\linewidth]{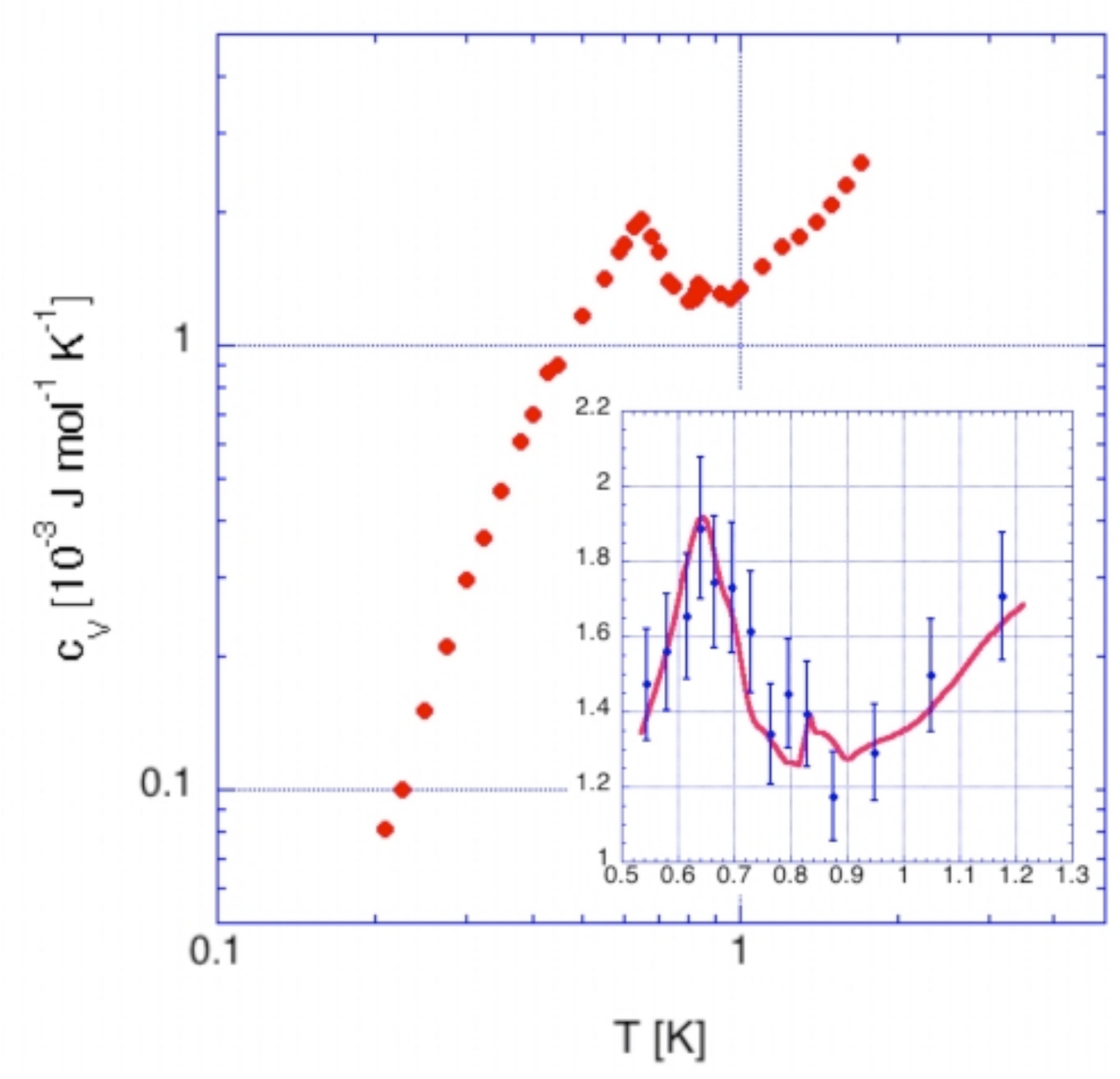}
       \caption{\it Al5056 specific heat: calorimetric measurements (5\% accuracy). The inset shows the calorimetric measurements interpolated by a polynomial (continuous line) and the independent $c_V$ determination based on the temperature increments at one end face of the bar (dots).}
       \label{cv}
\end{center}
\end{figure}

Samples of Al5056 obtained from the same production batch of the test mass have been used to characterize the material at very low temperatures. The measurement of the transition temperature to the $s$ state conducted using the mutual inductance method gives the value $T_c = 0.845 \pm 0.002\ \rm{K}$ and a total transition width of  about $0.1\ \rm{K}$. The smaller value of $T_c$ with respect to pure Al $(1.18\ \rm{K})$ could be ascribed to the presence of Mn impurities in the alloy. In fact, experimental studies on AlMn polycrystalline alloys have shown that $T_c$ was depressed down to $0.868\ \rm{K}$ and $0.652\ \rm{K}$ for Mn concentrations of 440 ppm and 900 ppm, respectively~\cite{smith}. Moreover, Al50XX alloys contain inclusions of the extremely  complex (MgAl) $\beta$ phase~\cite{powell} and the characterization of superconducting properties of the alloy $\beta-\rm{Al{_3}Mg{_2}}$ shows that $T_c=0.87\ \rm{K}$~\cite{bauer}.
Specific heat data for Al5056 are available in literature~\cite{coccia}, however, in order to completely characterize the production batch, we have performed  $c_V$ measurements above and below $T_c$ (Fig.~\ref{cv}) using  the calorimetric method of Ref. \cite{barucci}.
In the temperature interval $0.9\ \rm{K}\leq T \leq 1.5\ \rm{K}$ the fit of the data points, which have an accuracy of 5\%, to the function $c{_V}/T= \Gamma + B T^{2}$ gives the values $\Gamma = 1157 \pm 31\  \rm{erg\ cm{^{-3}}\ K{^{-2}}}$ for  the electronic specific heat coefficient  per unit volume in the $n$ state and $B = 0.14 \pm 0.01\  \rm{mJ\ mol{^{-1}}\ K{^{-4}}}$ for the lattice contribution. If the superconducting properties of Al5056 can be described by the BCS theory, then $H_c(0) \approx 2.42\  \Gamma{^{1/2}} T_c \approx 70\ \rm{Oe}$.
An independent check of the $c{_V}(T)$ behavior is obtained by the measurements at the bar end face of the temperature increments due to energy released by each beam pulse. The main features of the  $c{_V}(T)$ behavior, as obtained by the calorimetry, are well reproduced by this method (inset of Fig.~(\ref{cv})).

\begin{figure}[htbp]
\begin{center}
\includegraphics[width=1.0\linewidth]{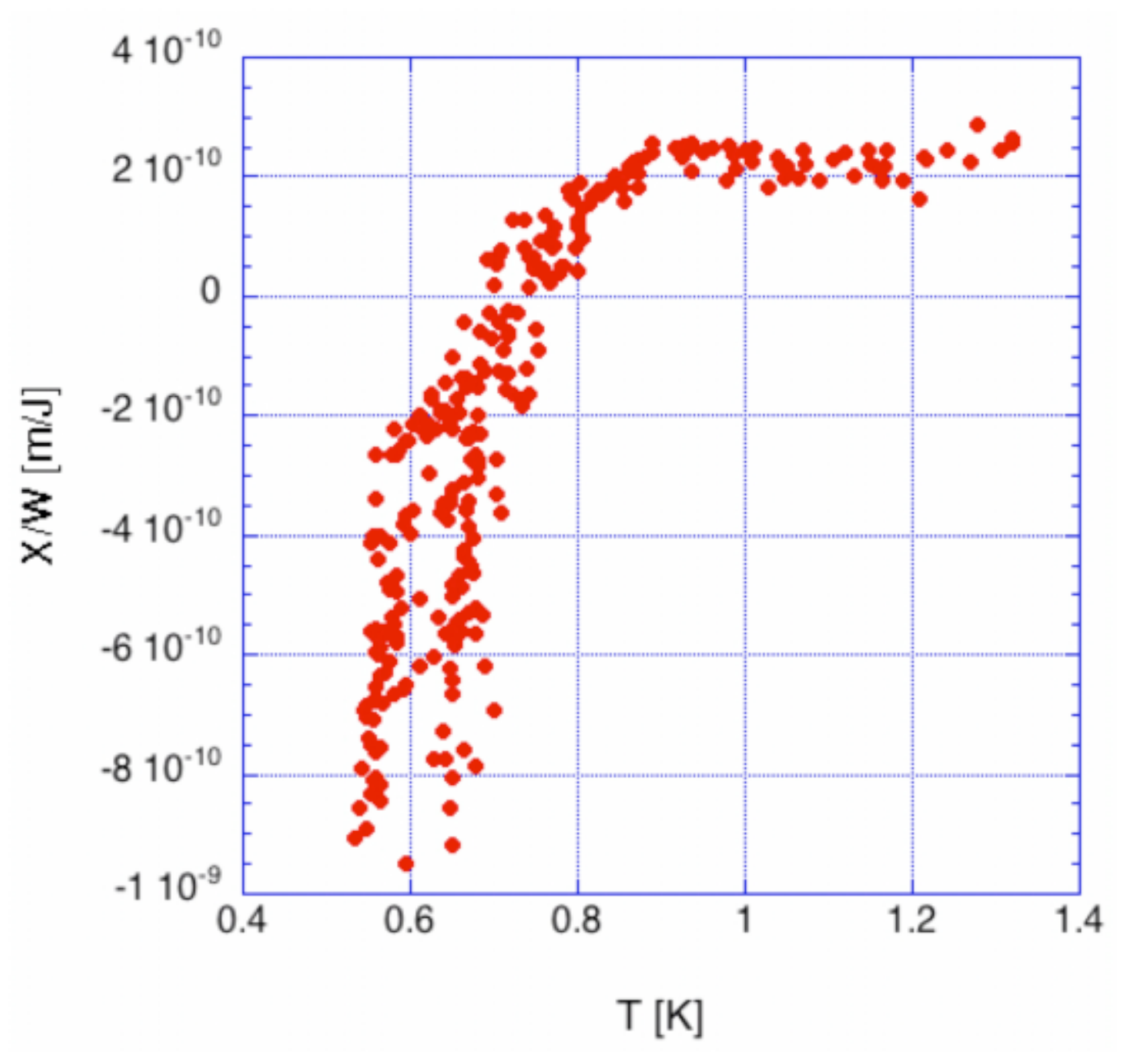}
      \caption{\it Measured  values of Amplitude $(X)$ normalized to the energy $(W)$ deposited in the bar per beam pulse  vs.  temperature $(T)$.}
        \label{alldata}
\end{center}
\end{figure}

\begin{figure}[htbp]
\begin{center}
\includegraphics[width=1.0\linewidth]{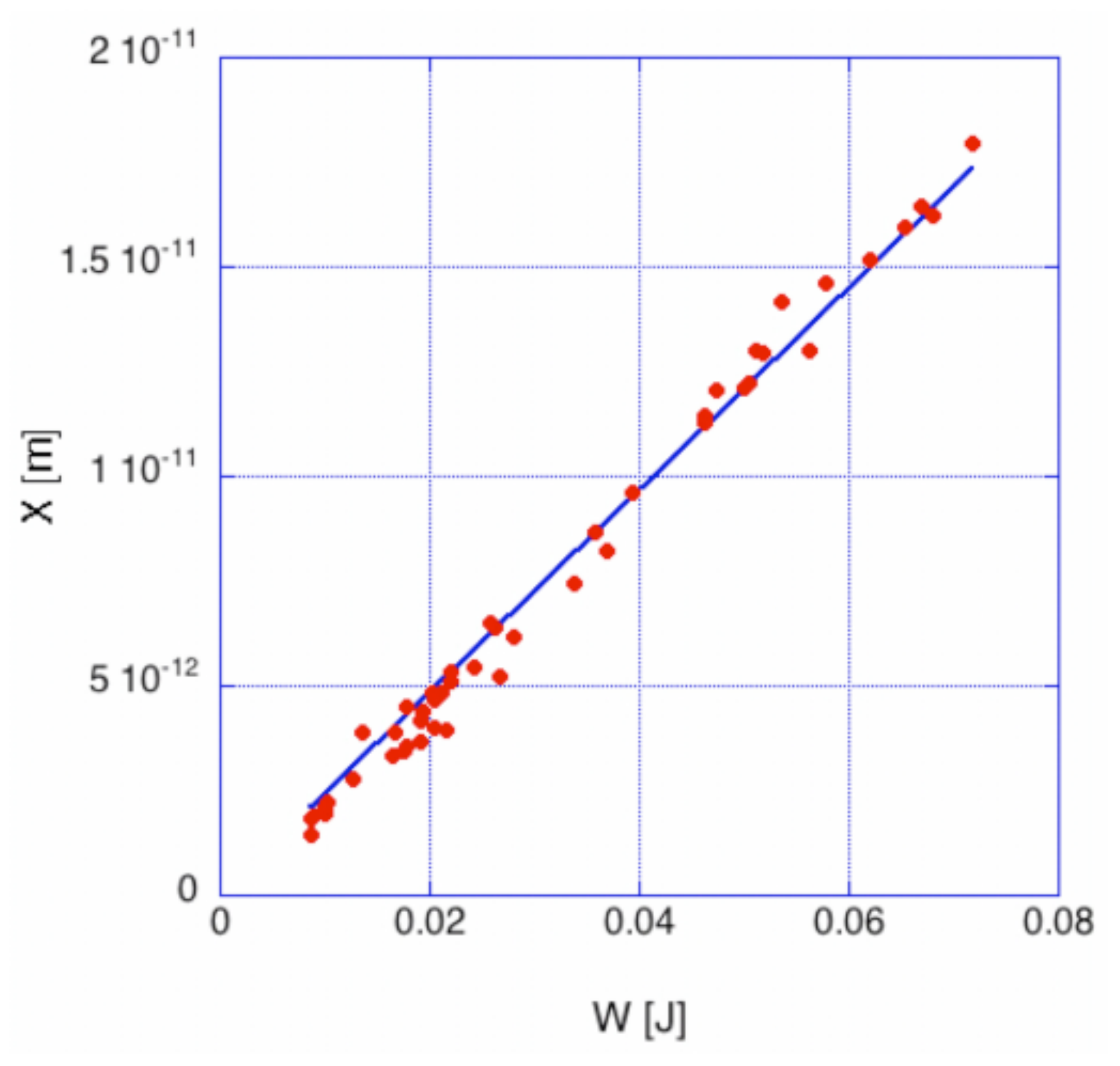}
        \caption{\it T$\geq 0.9\ K$ (n state); Measured values of Amplitude $(X)$ vs.  the energy deposited per beam pulse $(W)$. The slope of the fitted line is $ b=2.42\ 10^{-10}\ \rm{m/J}$.}
        \label{XoWnorm}
\end{center}
\end{figure}

The full set of Amplitude measurements $(X)$ normalized to the energy deposited per beam pulse $(W)$ in the explored temperature interval is shown in Fig.~\ref{alldata}. 
\noindent For $T \geq 0.9\  \rm{K}$, above $T_c$,  $X$ has a strict linear dependence on $W$, as expected from the relation (\ref{b0}). The linear fit  $X= bW$ (Fig.~\ref{XoWnorm}) gives $b=(2.42\pm 0.17)\ 10^{-10}\ \rm{m/J}$, where the error is determined by the quadrature of the beam monitor (3\%) and $\lambda$ determination (6\%) accuracies.
\noindent The onset at $T \sim 0.9\  \rm{K}$ and the behavior of the superconducting effects are shown in Fig. \ref{alldata}. As $T$ decreases, the normalized Amplitude becomes negative, indicating  that a
compression rather than an expansion is generated by the beam interaction in the
bulk.  Its absolute values is greater than  $b$, the normalized Amplitude value measured in the $n$ state. The increase in the absolute value of the Amplitude explains the effects seen in cosmic ray observations by NAUTILUS, when operated at $T=0.14\ \rm{K}$, as due to the conduction state of the material. 

\begin{figure}[htbp]
\begin{center}
\includegraphics[width=1.00\linewidth]{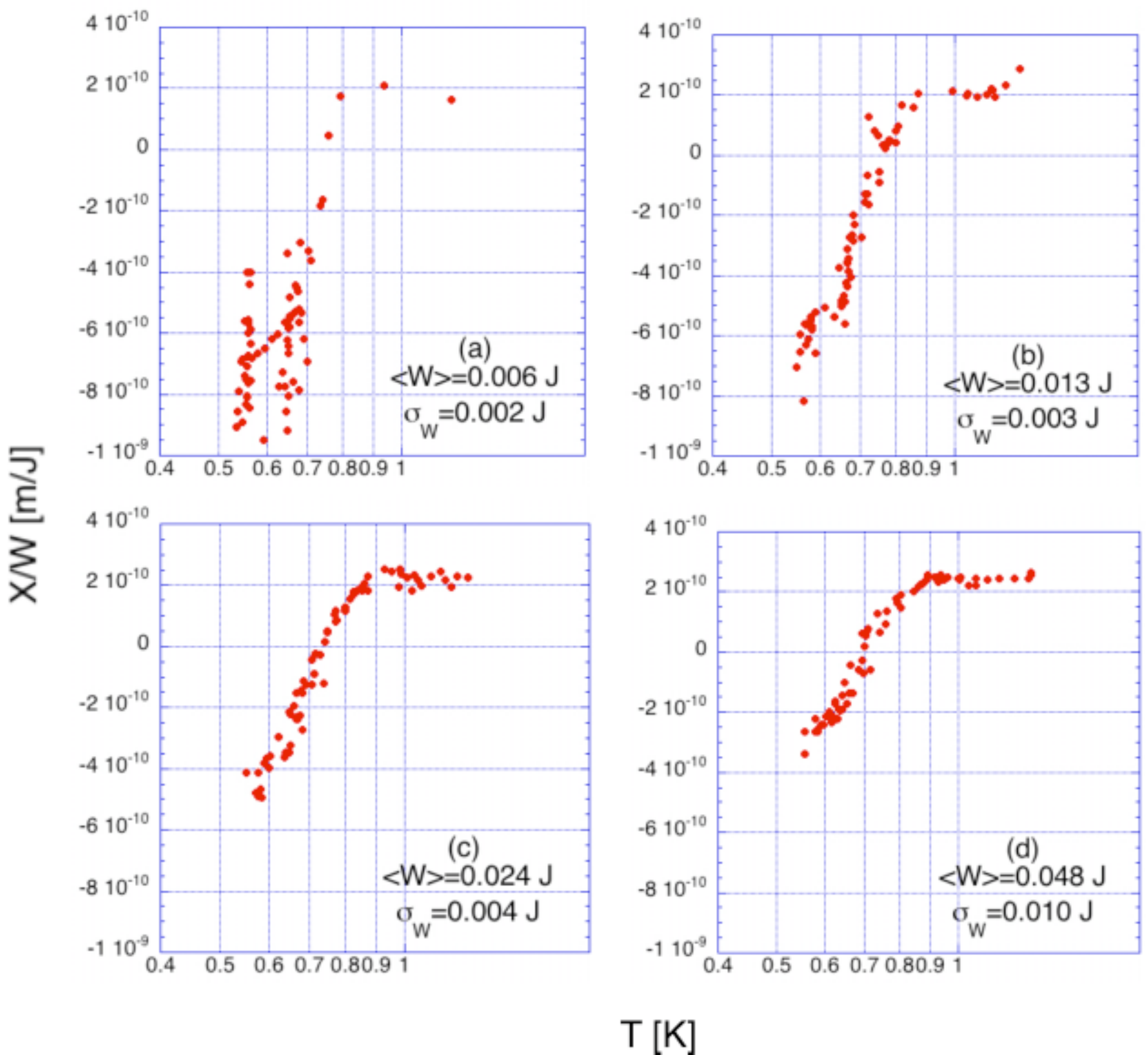}
       \caption{\it Data of Fig.~(\ref{alldata}) ordered in 4 bands of deposited energy. Each band is identified by $\langle W\rangle\pm\sigma_W$. A dependence of X/W on W can be seen in the $s$ state, e.g. at T=0.6 K.} 
        \label{wdep}
\end{center}
\end{figure}

\begin{figure}[htbp]
\begin{center}
\includegraphics[width=1.0\linewidth]{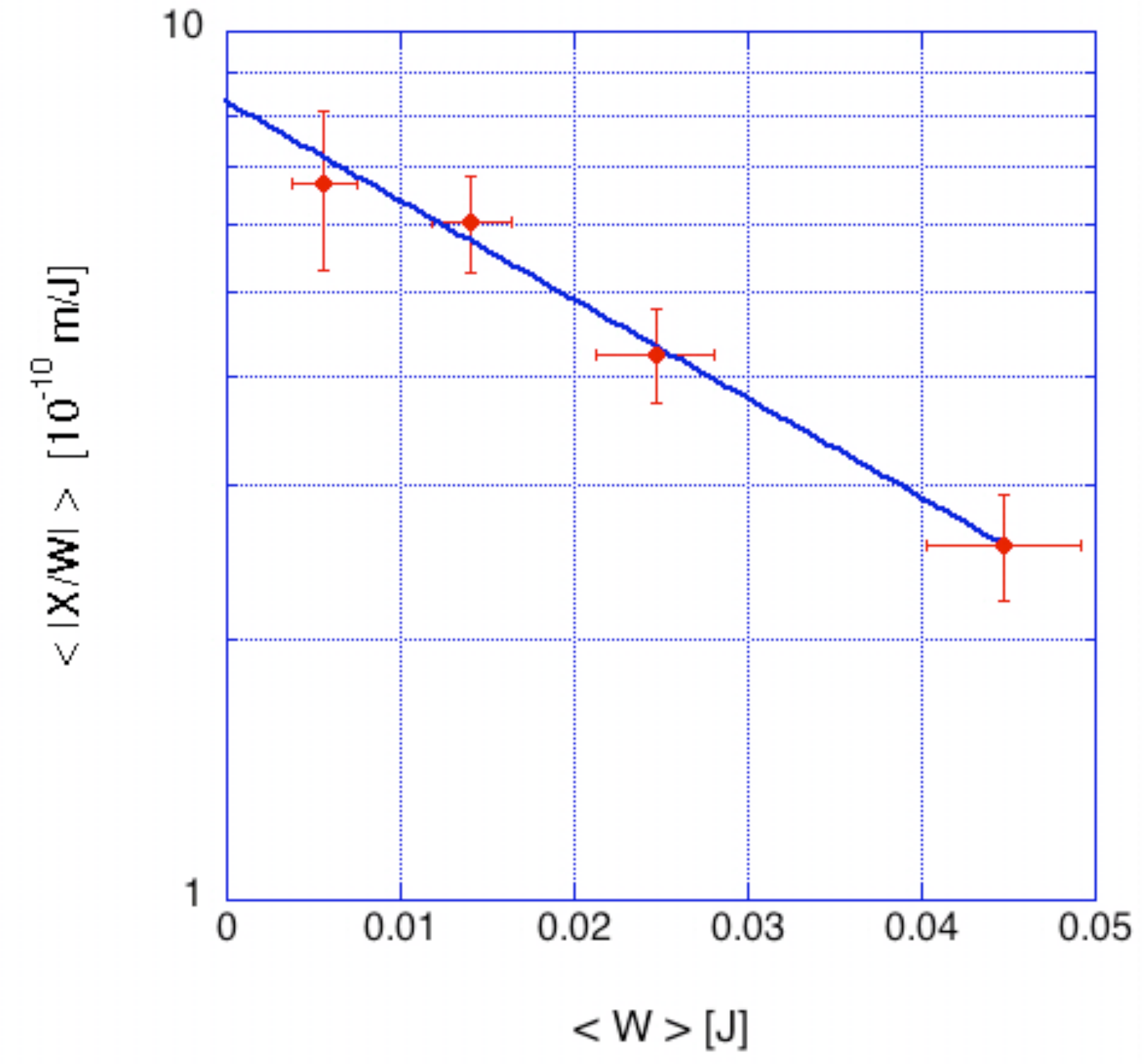}
     \caption{\it $0.55\leq T \leq 0.60\ K$; averages of the normalized absolute values of Amplitude $ \langle |X/W| \rangle $ vs.  the the average energy released per beam pulse $ \langle W \rangle $. The line represents the fit given by  $8.31\ 10^{-10}\ e^{-26.2\langle W\rangle}\ \rm{m/J}$.} 
    \label{extrap}
\end{center}
\end{figure}

\noindent Furthermore,  $X$ does not linearly depend on $W$ at fixed $T$, opposite to what has been observed \cite{bassan} in pure Nb  in the $s$ state. The dependence of $X/W$ on $W$ in the $s$ state is shown in Fig.~(\ref{wdep}) representing the data in four non-overlapping bands of $W$. 
This fact  has an impact on the quantification of the enhancement of the absolute value of the Amplitude below and far away from $T_c$. Fig. \ref{extrap} shows the averages of $|X/W|$ and $W$ in four  bins of data collected in the temperature interval ranging from 0.55  to 0.60~K, together  with the best fit given by the exponential $\langle |X/W| \rangle=(8.31\pm 2.88)10^{-10}\ e^{(-26.2\pm 6.3)\langle W\rangle}\ \rm{m/J}$. The average energy deposited by the cosmic rays interacting in the NAUTILUS antenna is in the order of $10^{-8}\ \rm{J}$ \cite{cosmici}, much lower than that released in our test mass by the beam pulse. We use  the ratio ${\mathcal F}=\langle |X/W| \rangle/b$  as a factor quantifying  the Amplitude enhancement in the $s$ state  with respect to the $n$ one. With reference to Fig.~(\ref{extrap}), we obtain ${\mathcal F}$=$3.4\pm 1.2$ by the extrapolation of the fitting exponential to $\langle W\rangle=10^{-8}\ \rm{J}$. A value of  ${\mathcal F} \sim 3.5$ is consistent with the cosmic ray observations made by NAUTILUS  in the $s$ state~\cite{cosmici}.

\section{Comparison with the model}
\label{comp}
The Amplitude ($X$) linearly depends on the deposited energy ($W$) in the model described in Section \ref{tam}, while a $X/W$ dependence  on $W$ is observed in the data. Therefore, we try to compare the model predictions to the data in the hypothesis that the linear dependence of $X$ on $W$ is attained at very low values of energy deposition.
The application of the model for the expected value $(X_{exp})$ computation of the Amplitude in the $s$ state requires the knowledge of 1) the thermophysical parameters $\alpha_n$ and $c_{V,n}$ of the material in order to evaluate $X_{therm}$ for the $n$ state below $T_c$ and 2) the dependence of  $H_c$ on $T$ and $P$ for calculating $X_{trans}$ via $H_c$ and its derivatives $\partial H_c/\partial T$ and $\partial H_c/\partial P$. The use of relations (\ref{b0}), (\ref{gru}),  (\ref{xth}) and (\ref{r}) allow us to write:
\begin{eqnarray}
 \lefteqn{\frac{X_{exp}}{W}= \frac{X_{therm}}{W} \left( 1 + {\mathcal R}\right)} \nonumber \\
 & =  & \frac{X_{therm}}{W} \left\{1 + \left[ \Lambda \frac{\Delta V}{V}+T \frac{\Delta \mathcal{S}}{V} \right] \left[{ \frac{\Delta\mathcal H}{V}}\right]^{-1} \right\} 
\label{exp}
\end{eqnarray}
\noindent with:
\begin{displaymath}
\Lambda = \frac{2\rho L (1+\epsilon)}{3\pi M\frac{X_{therm}}{W}}
\end{displaymath}
\noindent The requirement 1) cannot be fulfilled due to the lack of knowledge of $\alpha_n$ for Al5056 and we therefore assume that $X_{therm}/W=b$ also in the temperature interval $0.5\ \rm{K} \lesssim T \leq T_c$, due to the fact that $\gamma_n$, in this interval, is expected to have almost the same value as that assumed at slightly higher temperatures. In relation to requirement 2), we derive $\partial H_c/\partial T$ at $T<T_{c}$ from the $H_c$ parabolic dependence  on $t$, assuming that the unknown dependence of $\partial H_c/\partial P$ on $t$ at $P=0$ for Al5056 is equal to that of pure Al. Under this hypothesis, $\partial H_c/\partial P$ can be deduced by interpolating the tabulated values of $H_c$ as a function of $T$ and $P$ contained in Ref.~\cite{harris}. Inserting numerical values in relation (\ref{exp}) gives an average of 
$\langle X_{exp}/W \rangle=(-18\pm 1)10^{-10}\ \rm{m/J}$ in the interval $0.55 \leq T\leq 0.6\ \rm{K}$, where the error does not include systematic contributions deriving from the assumptions made. This is to be compared to $\langle X/W \rangle=(-8.3\pm 2.8)10^{-10}\ \rm{m/J}$, obtained at $W=10^{-8}\ \rm{J}$ from the measurements in the same temperature range (see Fig.(\ref{extrap})). This  discrepancy can be ascribed to the fact that the model, as mentioned by the authors of Ref.~\cite{deru}, considers superconducting effects in pure materials, while the intrinsic properties of an alloy could determine additional contributions to the expected values of the oscillation amplitudes. Indeed, the model has given a satisfactory description of the behavior of the experimental data collected with a pure Nb test mass in the $s$ state~\cite{bassan}. 

The calculation of $X_{exp}/W$ for the $s$ state, in the framework of  the alternative scenario described in Section~\ref{tam}, requires the knowledge of $\alpha_s(=\alpha^e_s+\alpha^r$,  where $e$ and $r$ refer to the electronic and lattice contributions, respectively) and $c_{V,s}$,  the former being unknown and the latter measured.  The relation~(\ref{gru}) gives $\alpha^e_s=\rho c_{V,s}{^e} \gamma^e_s/(3 K_T)$ and $\alpha^r=\rho c_{V}{^r} \gamma^r/(3 K_T)$. Again, for lack of better knowledge, we presume the values of  the  Gr\"{u}neisen parameters and of $K_T$ for AL5056  to be similar to those of pure Al. Thus, we use $\gamma^r=2.6$~\cite{barron} in the limit $t\rightarrow 0$, $\gamma^e_s=-11.5\pm 1.0$ at $t\sim 0.7$~\cite{marini} and $K_T=79.4\ 10^9\ \rm{N/m^2}$ near T=0 \cite{kamm}.
The insertion of these values in relation (\ref{xth}) gives $\langle X_{exp}/W \rangle=(-11\pm 1)10^{-10}\ \rm{m/J}$ in the same temperature interval as in the first scenario. As previousy, the error is determined only by  propagating  the errors on the quantities in the right side of relation (\ref{xth}). The systematic uncertainties introduced in the calculations of the expected values in both scenarios do not allow us to individuate the one that better agrees with the experimental data. Moreover, it is interesting to derive the predictions at temperature values close to the lower limits of $T$ in the $H{_c}(P,T)$ tabulation  of Ref.~\cite{harris}: $X_{exp}/W =(-28\pm 1)10^{-10}\ \rm{m/J}$ at $T= 0.3\ \rm{K}$ in the first scenario and $X_{exp}/W =(-28\pm 16)10^{-10}\ \rm{m/J}$ in the alternative scenario, where for the latter the large error is due to the uncertainty on  $\gamma^e_s$ at this temperature.

Finally, dissipative effects, which can be inherent in this alloy and due to the flux line motion with consequent entropy transport, could play a role in the  Amplitude observed values. A clue in this direction lies in the fact that Amplitude is not linearly dependent on $W$ in the $s$ state at fixed $T$. 
 
\section{Conclusions}
 \label{conc}
The measurements performed on an  Al5056 suspended bar, hit by an electron beam and operated at temperatures above and below $T_c$, have shown that in the $s$ state, the amplitude of the fundamental mode of the bar is enhanced  with respect to the $n$ state by a factor $\sim$ 3.5 at $T\sim 0.5\ \rm{K}$. This factor is consistent with the observations made by NAUTILUS on cosmic rays at $T=0.14\ \rm{K}$. The amplitude change in the $s$ state, following an energetic particle interaction, is due to the superconducting properties of the material. The absolute value of the normalized Amplitude is enhanced in the Al alloy and reduced in Nb. Incomplete knowledge of the  involved thermophysical and thermodynamic parameters does not allow a full assessment of the model describing the underlying physical process in the $s$ state. The effects due to cosmic ray interactions could be an important source of noise in  future gw acoustic and interferometric detectors of improved sensitivities and a complete characterization of the thermo-acoustic effects in  the test masses operated in the $s$ state should be performed by direct measurements of the type shown in this letter.

\section*{Acknowledgements}
We thank Messrs.~ F.~Campolungo, M.~Iannarelli and R.~Lenci, which helped with  the experiment setup. 

\noindent This work is partially supported by the EU Project ILIAS (RII3-CT-2004-506222).




\begin{thebibliography}{99}
\bibitem{cgs} We keep the practical cgs unit for the magnetic field, as used by the authors of the cited articles,and we convert the density of the magnetic energy to SI units.
\end{thebibliography}
\end{document}